\documentstyle[12pt,boxedminipage,epsfig,wrapfig,floatflt,shadow]{article}

\begin{document}

\title{
Assessing Student Expertise in Introductory Physics with Isomorphic Problems, 
Part I: Performance on Non-intuitive Problem Pair from Introductory Physics} 
\author{Chandralekha Singh\\ Department of Physics and Astronomy\\ University of Pittsburgh, Pittsburgh, PA, 15260}
\date{ }

\maketitle

\vspace*{-.2in}

\begin{abstract}

Investigations related to expertise in problem solving and ability to transfer learning
from one context to another are important for developing strategies to help
students perform more expert-like tasks.
Here we analyze written responses to a pair of non-intuitive isomorphic problems given to 
introductory physics students and discussions with a subset of students about them.
Students were asked to explain their reasoning for their written responses.
We call the paired problems isomorphic because they require the same physics principle to solve them.
However, the initial conditions are different and 
the frictional force is responsible for increasing the linear speed of an object in one of
the problems while it is responsible for decreasing the linear speed in the other problem.
We categorize student responses and evaluate student performance within the context of their evolving expertise.
We compare and contrast the patterns of student categorization for the two isomorphic problems.
We discuss why certain incorrect responses were better than others and shed light on the evolution of students' expertise.
We compare the performance of students who worked on both isomorphic problems with those
who worked only on one of the problems to understand whether students recognized their underlying similarity
and whether isomorphic pairs gave students additional insight in solving each problem.

\end{abstract}

\vspace*{-.2in}
\section{Introduction}
\vspace*{-.2in}

Developing expertise in problem solving constitutes a major goal of most physics courses~\cite{newell,fred,maloney,mestre,alan}.
Problem solving can be defined as any purposeful activity where one is presented with a 
novel situation and devises and performs a sequence of steps to achieve a set goal~\cite{chi}.  
Both knowledge and experience are required to solve the problem efficiently and effectively.  
Genuine problem solving is not algorithmic; it is heuristic. 
There are several stages involved in effective problem solving, including initial qualitative analysis, planning,
assessment, and reflection upon the problem solving process in addition to the implementation stage~\cite{schoenfeld}.
The problem solver must make judicious decisions to reach the goal in a reasonable amount of time.
Given a problem, the range of potential solution trajectories that different people may follow to achieve the goal can be
called the problem space~\cite{newell}. For each problem, the problem space is very large and based upon one's expertise, people
may traverse very different paths in this space which can analogically be visualized as a maze-like structure.

Several studies have focused on investigating the differences between the problem solving strategies employed by experts
and novices~\cite{hardiman,chi3,jong,larkin,intuition}. 
These studies suggest that a crucial difference between the problem solving capabilities of experts and beginners 
lies in both the level and complexity with which knowledge is represented and rules are applied.  
Expert knowledge is organized hierarchically in pyramid-like knowledge structures where the most fundamental concepts are at the top
of the hierarchy followed by the ancillary concepts. 
Experts view physical situations at a much more abstract level than novices. For example, experts in physics consider
a problem involving angular speed of a spinning skater moving her arms close or far from her body very similar
to the problem related to the change in the speed of a neutron star collapsing under its own gravitational force. 
Rather than focusing on the ``surface" features of the two problems: a spinning skater in one case and rotating neutron star in
the other case which appear very different,
experts focus on ``deep" features based upon the abstract physics principle; the fact that there are no external torques on 
the relevant system in each case so that the angular momentum is conserved. 
Novices on the other hand often focus on these surface features, may get distracted by irrelevant details, and may not
see the inherent similarity of the two problems.  
Two classic studies of problem categorization in introductory mechanics problems indicate that novices categorize problems according
to the objects of the problems, regardless of the physical principles required for solving them~\cite{hardiman,chi3}.
For example, novices deemed problems similar if they involved inclined planes, or pulleys, or springs, as opposed to whether they
could be solved by applying Newton's laws or conservation of energy.
In contrast, physics experts categorize problems based on physics principles, not the problems' surface similarity~\cite{hardiman,chi3}.
Experts' knowledge representation and organization along with their superior problem solving strategies
help them narrow the problem-space without cognitive overload and retrieve relevant knowledge efficiently from memory~\cite{schema,rep,zhang}.
Although expertise studies usually classify individuals either as an expert or a novice, people's expertise in a particular
domain spans a large spectrum in which novices and ``adaptive" experts are at the two extremes~\cite{hatano}.

Simon and Hayes defined two problems as isomorphic if they have the same structure to their problem space~\cite{hayes1,hayes2}. 
They were the first to analyze why one problem in an isomorphic problem pair (IPP) may be more difficult than the other
using their model of problem solving~\cite{hayes1,hayes2}.
Cognitive theory suggests that the context in which knowledge is acquired and the way it is stored in memory
has important implications for whether cues in a problem statement will trigger a recall of the 
relevant concepts~\cite{simon,context,diver}. 
Depending upon the context, the problem space for the problems in an IPP may be such that 
one problem may trigger the recall of relevant concepts from memory while the other problem may not. 
The famous ``Tower of Hanoi problem" is isomorphic to the ``cannibal and the missionary problem"~\cite{hayes1,hayes2,mission}.
Research shows that the Tower of Hanoi problem in this IPP is more difficult than the latter~\cite{mission}. 
Despite the same underlying features of these problems, the problem solvers, in general,
traverse very different trajectories in the problem space and use different knowledge resources while solving the two isomorphic problems.

The isomorphic pairs chosen by Simon and Hayes shared ``deep" features but had very different surface features involving pegs and disks of
varying radii in the Tower of Hanoi problem, and cannibals, missionaries, river and boats in the other. 
Here, we will define a pair of problems as an isomorphic problem pair (IPP) if they require the same physics principle to solve them.
The similarity of the problems in an IPP can span a broad spectrum. Isomorphism between problems has been observed in studies about students'
conceptions, e.g., in the context of changes of reference~\cite{euro}. Very closely related IPPs may involve problems in
which the situation presented is the same but some parameters are varied, e.g., two similar projectile problems with different initial
speed and/or angle of launch. One level of difficulty with regard to discerning their similarity
can be introduced by changing the context of the pair problems slightly.
For example, two isomorphic problems about projectiles can be about a person kicking a football or throwing stones
from a cliff. Depending upon an individual's level of expertise, the person may or may not discern the similarity of these problems
completely. Another level of difficulty can be introduced, e.g., by making one problem in the IPP quantitative and
one qualitative. A high level of complexity can be introduced by making the surface features of the problems very different 
as in the problem pair chosen by Simon and Hayes or by introducing distracting features into one of the problems.

Although most educational and cognitive researchers are ultimately striving to gain insight into how to
improve learning, a survey of the previous literature shows that the analyses regarding the interpretation of student responses differ.
The analysis can span a wide spectrum ranging from a focus on differentiating between experts and beginning students to showing 
the similarities in their responses.
In some cases the analyses may be complimentary even if the focus is different but in other cases researchers may express
diverging views on what the students' responses suggest about their cognition and their use of 
problem solving and meta-cognitive strategies.
For example, the following problem depicted in Figure 1 was given to an expert and a beginning student by Larkin et. al.~\cite{larkin2}: 
``What constant horizontal force $F$ must be applied to
the large cart (of mass M) so that the smaller carts (masses $m_1$ and $m_2$) do not move relative to the large cart? Neglect friction."
Larkin et. al. report that in response to this question, the expert invoked the idea of accelerating reference frame and pseudoforce to 
justify why $F$ will prevent
smaller masses from moving relative to the large cart while the student said that the wind must push on $m_1$ so that $m_2$ does not fall.
Larkin et. al.~\cite{larkin2} argue that although the expert used a sloppy language while invoking pseudoforce, the expert analysis
was deep while the novice analysis was superficial. They suggest that most experts
will disagree that both the expert and the novice performed equally deep analyses. In particular, if the interviewer had qualified
her problem statement by saying that ``neglect friction" means ``neglect friction and the effect of air", the novice would have had
difficulty in proceeding while the expert would have still succeeded.
On the other hand, Smith et. al.~\cite{smith} argue that both the expert and novice response show equally deep analysis of the problem citing: 
``It is hard to see how the
expert's pseudoforce, as characterized by Larkin, is any more abstract than the novice's wind. Both
the wind and the whiplash force are constructions inferred from their effect. Both expert and novice also rapidly simplify and reformulate
the problem, producing a deep analysis of the situation." They further add~\cite{smith}: 
``Both their solutions represent selection of deep features that cut to the
physical heart of the problem."

\vspace*{-.2in}
\section{Goals}
\vspace*{-.2in}

Here we analyze the performance of college introductory physics students from two different courses on a non-intuitive 
isomorphic problem pair (IPP). Although both isomorphic problems involved rotational and rolling motion, the initial conditions
were very different. In one problem, friction
increased the linear speed of an object until it started to roll, while in the other problem, friction decreased the linear speed of
the object until the rolling condition was satisfied. 
The first goal was to categorize student responses and evaluate student performance within the context of their evolving expertise.
Here, the phrase ``evolving expertise" refers to the fact that that expertise in a particular domain can vary widely and all introductory
physics students may not necessarily be novices in the sense that they will not group all the inclined plane problems in physics in one category if asked to categorize problems. 
Many introductory students
may have developed sufficient knowledge and skills that their expertise in solving introductory physics problems may have evolved
to an intermediate or 
even advanced level. In this sense, the phrase ``evolving expertise" in this paper
has not been used to imply a dynamic connotation, something that could occur during the problem solving process.

A second goal was to compare and contrast the patterns of student categorization for the two isomorphic problems and explore these differences
in light of the depth of reasoning and analysis performed by students rather than focusing on the correctness of their answers.
We analyze why certain incorrect responses are better than others and highlight the evolution of students' expertise.
Our analysis of student responses and interviews suggests that the approach taken by some students was superior to others as some
of them critically evaluated the problems given to them and weighed different options carefully.
Some student responses show signs of their evolving expertise and their resemblance to experts in some respects.

A third goal was to compare the performance of students who worked on both problems in the IPP with those
who worked only on one to understand if working on one of the isomorphic problems
affected the performance on the other problem in that pair. 
We analyze why students may invoke different knowledge resources in different contexts to solve the isomorphic problems
with the same underlying physics principle and whether these knowledge resources are different from those invoked by experts.
These issues will then be explored further via a range of contexts in a companion paper.

\vspace*{-.2in}
\section{Methodology}
\vspace*{-.2in}

In this investigation, we developed a pair of isomorphic introductory physics problems related to rotational and rolling motion.  
Students in two calculus-based introductory physics courses
were given the IPP related to rotational and rolling motion in the free-response format as opposed to the multiple-choice format. 
The IPP and the solutions of the problems are described in Appendix 1. 
The isomorphism in these problems is at the level of the
physics principle involved. However, the initial conditions are very different in the two problems. In one problem, the force
of friction increases the linear speed of the object until it begins to roll, while friction decreases the linear speed of
the object in the other case before the rolling condition is met.

These problems were given to students after traditional instruction of relevant concepts in lecture format and after students
had the opportunity to work on homework problems from the relevant chapter.
Students who were given these problems had class discussions and homework problems about situations in which the frictional
force assists in maintaining the motion of an object. For example, there was a discussion of why a crate on the floor of a truck or
a cup on the airplane-tray in front of your seat does not fall (get left behind) when the truck or the plane accelerates forward. 
There was also a class demonstration and discussion of why a glass full of water does not fall when a table cloth is pulled with a 
jerk from underneath it but it falls if the cloth is pulled slowly. There was no formal laboratory component to these courses
but students did exploration homework problems each week which were closely tied to lecture demonstrations~\cite{pec}.
The condition for rolling without slipping was also discussed extensively.

We first administered these problems in the form of a recitation quiz to a calculus-based introductory physics class.
Out of a class of 137 students, 67 solved Problem 1 and 70 solved Problem 2.
In another calculus-based introductory physics class, 49 students were given both problems of the IPP
in a recitation quiz. 
In addition to asking students to explain their reasoning, we discussed their intuition and approach individually with 
several student volunteers to better understand how they had interpreted and answered the problems. 
An additional open-ended question was given to
those students who answered both problems of the IPP: they were asked whether the two problems they
solved are similar or different and why.

Below, we categorize student responses to each of the problems and discuss what we learned from the 
responses about students' levels of expertise even if the responses are not completely correct. 
We also compare responses for the case where students were given only one of the problems of the IPP
with the case where they were given both problems of the IPP simultaneously.

\vspace*{-.2in}
\section{Results}
\vspace*{-.2in}

The problem statements did not explicitly specify whether students should solve the problems qualitatively or quantitatively and
students could have solved it either way. In case a student solved a problem quantitatively, the student could then have
made qualitative inferences based upon the quantitative solution to interpret that the final speed of the object before the
rolling condition is satisfied is independent of friction for both problems. None of the students in this study chose to solve the problem 
quantitatively and their problem solutions involved conceptual reasoning.

As discussed earlier, here we 
analyze students' responses based not simply upon their absolute correctness, but
the extent to which they resemble expert responses and reflect students' evolving expertise during a transitional period.
As noted earlier, students' partially correct responses 
can be interpreted differently by researchers.
Here we explore the extent to which students are capable of performing 
problem analysis similar to the type that is expected from experts.
The student responses to Problems (1) and (2) can be classified in five broad categories:
\begin{itemize}
\item Category 1: Friction will act in a direction opposite to the velocity and slow the object down. Therefore, larger $\mu$ implies smaller $v_f$. 
\item Category 2: Since the frictional force is responsible for making the object roll, higher $\mu$ should imply higher $v_f$. 
\item Category 3: Since larger friction implies shorter slipping time, the $v_f$ will be larger in this case. 
\item Category 4: $v_f$ will be independent of $\mu$. Although this is the correct response, 
as discussed later, we consider the response correct only if a correct reasoning was provided. 
\item Category 5: Responses which did not appropriately address the question that was asked or did not fall in any other categories.
\end{itemize}

\subsection{Student responses to Problem (1)}

We first analyze the response of students who were only asked to respond to problem (1).
Column 2 of Table (1) shows the fraction of students in each of the categories above.

Responses of $43\%$ of the students were in Category (1). These students
thought that friction will reduce the linear velocity because the two must oppose each other. 
They often believed that the problem was relatively easy because they felt that the friction on the floor
can only decrease the speed of the wheel before it starts rolling.
Individual discussions show that several students in this category did not differentiate between the linear and angular speed. When they were 
explicitly asked about whether there was a horizontal speed at the time the wheel hit the floor, some started to worry that they were confusing
the linear and angular speeds. 
Even after this realization, many in this group were convinced that friction could not increase the linear speed. 
Some hypothesized that there must be a force in the direction of motion in addition 
to the retarding frictional force to ensure that there was a linear speed when the wheel hits the ground. 
Others in Category (1) who did not mix up the linear and angular speed continued to support their original response. 
Some assumed that the wheel will develop a linear speed as soon as it hits the ground.
When asked explicitly about what will cause it to develop the linear speed, some
noted that the impact will produce a linear speed as soon as the wheel hits the ground, others said
that there has to be a force in the direction of motion without actually identifying it, 
and a few admitted that they could not at the moment think of a good reason for it. 
Some students confused the vertical speed of the falling wheel with its horizontal speed.
Incidentally, several students in Category (1) drew diagrams with a force in the direction of velocity in addition
to drawing a frictional force acting in the opposite direction.
The following responses from Category (1) show additional difficulties. In some cases, we explicitly point out the difficulty that is
inferred from the student responses:

\begin{itemize}
{\setlength{\rightmargin}{\leftmargin}}
\itemsep -7pt

\item{\it $v_f$ will be larger if the wheel falls on ice because ice is almost frictionless so it will roll faster with less friction holding it back.}

\item{\it The translation effectiveness of wheel depends upon friction.  There is energy needed to continue motion.
Rough surface would decelerate it at a faster rate and give smaller $v_f$.}

\item{\it Larger $\mu$ will slow the object more because large friction with consideration to the normal force is working against the forward velocity.} 

\item{\it Friction acts in the direction opposite to the force moving the wheel in the horizontal direction so it makes the force less than what it were
on frictionless surface. More $\mu$ will slow it more.} [notion that there must be a force in the direction of motion]

\item{\it $v_f$  will be larger on ice because ice provides very little friction so that the wheel doesn't have to push itself with a lot of force
to keep moving.} [notion that the wheel has to push itself to keep moving]

\item{\it $v_f$ will be larger on ice because smaller $\mu$ will allow the wheel to retain more of its original angular momentum.}

\end{itemize}

Twenty-seven percent of students were in Category (2).
 They correctly knew that the directions of the kinetic frictional force and $v_f$ 
are the same. Analysis of written responses and discussions with individual students suggest that students in this category analyzed the problem
more critically and deeply than those in Category (1). 
Unlike the assumption made by students in Category (1) that the frictional force must act
in the direction opposite to the linear velocity of the wheel, these students realized that the frictional force was responsible for imparting
a linear velocity to the wheel and for getting the wheel rolling (the wheel initially only had an angular velocity).
This type of reasoning is key to solving the problem correctly and shows a sophisticated reasoning similar to those of experts.
Students in Category (1) were exploring the region of the problem space that led to a dead end and could
not have taken them closer to the correct problem solution. On the other hand, students in Category (2)
were headed in the right direction. Their analysis is incomplete but not totally incorrect because they evaluated
the role of friction correctly but did not take into account the amount of energy dissipated in the form of heat 
before the wheel started rolling.
The following are sample responses from this category:

\begin{itemize}
{\setlength{\rightmargin}{\leftmargin}}
\itemsep -7pt
\item{{\it $v_f$ will be larger if it fell on a rough surface because $f$} [friction] {\it is responsible 
for forward motion.}}

\item{\it $v_f$ will be larger when $\mu$ is larger. When the road is rough, cars are able to go over 200 mph
without slipping, on ice the car would just slide.}

\item{\it $v_f$ will be larger on a rough surface because the increase in friction will allow the angular velocity to be converted into tangential
velocity while gripping the surface.}

\item{\it A wheel ``pushes" against the ground to move horizontally. According to Newton's 3rd law the ground pushes back. Higher $\mu$ makes
the wheel grip the ground better and $v_f$ is higher.}

\item{\it When the wheel hits a surface that provides a larger $\mu$, the spinning will catch the surface and the wheel will move faster.}

\item{\it Larger $\mu$ implies larger $v_f$ because the wheel will get more traction from the surface and cover more ground in the linear
direction.}
\end{itemize}

Nine percent of the student responses fell in Category (3). These students correctly noted that a larger $\mu$ would imply that
the wheel will reach the final steady state in less time. Similar to the responses of students in Category (2), these students also
analyzed the problem deeply.
Although they did not solve the problem
correctly, they correctly noted that the frictional force will assist in increasing the linear velocity.
This is in contrast to the superficial responses of many students in Category (1) who asserted that 
friction can only decrease the linear velocity so the larger friction must result in a smaller $v_f$.
The analysis of Category (3) students is at least partially correct because a larger frictional force will definitely
cause the wheel to ``lock" in faster and begin to roll more quickly. 
What these students overlooked was that the larger frictional force would also lead
to a higher power dissipation. If they had combined their partially correct analysis with the fact that the energy dissipated per unit
time while the wheel is slipping is more for higher friction, they may have navigated through the problem space successfully
and reached the finish line.
The following are sample responses from Category (3):

\begin{itemize}
{\setlength{\rightmargin}{\leftmargin}}
\itemsep -7pt
\item{\it When the spinning wheel is dropped, the quicker it gets rolling, the more speed it will have. $\mu$ must be large for the wheel to
``catch" quickly so that much of the speed of the spinning wheel is not lost.}

\item{\it If the wheel spins a lot before rolling it loses its momentum whereas if a rough surface provides friction it is able to obtain a
greater velocity quickly.}

\item{\it Contact with a surface with a larger $\mu$ will cause the wheel to slip for a shorter time and allow more energy to be put into
$v_f$ making it faster.}
\end{itemize}

In fact, problem (1) was also given to twenty college physics faculty (experts)~\cite{intuition}. 
The problem is one for which professors have very little
physical intuition and it puts them in a situation similar to the students where they have to think
``on-their-feet" to construct a solution rather than invoking ``compiled" knowledge from memory~\cite{simon}.
Although professors would have solved the problem without the time constraint,
our goal was to elicit the thought-processes and problem solving strategies of experts as they venture into
solving a non-intuitive problem. In quizzes and examinations, students often work under a similar time constraint. 
The problem given had two important variables that were inversely related to $v_f$: the force of friction and the time to start rolling.  
Most professors admitted that they did not have much intuition about how the final speed $v_f$ should depend on the 
coefficient of friction, $\mu$.  
Although they initially employed superior problem solving strategies, they
had great difficulty in thinking about the effect of both important parameters in the problem similar to students in Categories (2) and (3). 
Seventeen out of the twenty professors concentrated almost exclusively on one of the two 
essential features of the problem, either the frictional force or the time to start rolling. 
Those who focused on the time to roll often noted that a high friction would lead to quicker rolling so less energy 
will be dissipated in that case and $v_f$ will be larger. Those who focused on friction and did not account for the time to 
roll, typically concluded that a high friction would lead to more energy dissipation and hence a smaller $v_f$.  

Comparison of student and expert responses shows that students in Categories (2) and (3) should not be classified as beginners on 
the expertise scale. 
The fact that, in an unfamiliar situation, even experts struggle to focus on more than one important aspect of the problem suggests
that we should not expect students in the introductory physics courses to have focused on both aspects of the problem to solve
it correctly. The responses in Categories (2) and (3) resemble the responses of the experts and point to students' evolving expertise.

Students in Category (4) ($10.5\%$) said that $v_f$ is independent of $\mu$. Although this response appears to be correct on the surface, 
all but one student
provided incorrect reasoning. Students who provided incorrect reasoning often focused on the motion after the wheel started to roll rather
than the effect of friction and the dissipation of energy in the form of heat during the slipping process.
The following are sample responses from Category (4). The first reasoning is qualitatively correct whereas the second example shows
an incorrect reasoning:

\begin{itemize}
{\setlength{\rightmargin}{\leftmargin}}
\itemsep -7pt
\item{\it $v_f$ depends on how much energy is lost to heat due to friction.  On a rough surface, heat will be dissipated quickly, whereas on
an icy surface the same heat will be lost over a longer time. So $v_f$ will be same for all $\mu$.}

\item{\it $v_f$ is independent of $\mu$ because $v_{cm}=r \omega$ by the definition of rolling depends only on $r$ and $\omega$.}

\end{itemize}

Students in Category (5) ($10.5\%$) provided responses that were unclear. It appeared that the students either did not read the question
carefully or did not analyze and formulate their responses carefully.
The following type of response is inconsistent with the problem given (Category (5)):

\begin{itemize}
{\setlength{\rightmargin}{\leftmargin}}
\itemsep -7pt
\item{\it $v_f$ will be larger while the wheel is slipping and smaller when it grips.}
\end{itemize}

\subsection{Student responses to Problem (2)}

Seventy students were asked to solve only problem (2).
Although the explanation for the independence of $v_f$ on $\mu$ for Problem (2) is the same as that for Problem (1),
there is a crucial difference in the surface features of the two problems because the initial conditions are very different
for the two problems. There is a non-zero initial linear speed of
the pool ball when it is struck as opposed to a non-zero initial angular speed of the wheel when it is dropped on the floor.
This implies that friction increases the linear speed in one case and decreases it in the other case before each object
starts to roll.
This difference led to a different distribution of student responses which can be classified into four of the categories
used for problem (1) as shown in Table (1).

A comparison with the student responses to Problem (1) shows that the responses in Category (1) almost doubled for Problem (2).
In particular, in the context of the pool ball which had the initial linear speed $v_0\ne 0$, $76\%$ of the students believed
that a higher frictional force will make $v_f$ smaller when the rolling begins. 
This shift is due to the fact that in the wheel problem, friction helps in increasing the linear 
speed and in the pool ball problem it decreases it.
This shift and individual discussions with students suggest that the spinning wheel dropped on the floor forced many students to think about
why the wheel will pick up linear speed when it falls on the floor and eventually starts to roll. 
In the pool ball problem, the 
idea that ``a higher frictional force must decrease $v_0$ more and lead to a smaller $v_f$" sounded robust to many students as 
can be inferred from the following representative responses from Category (1):

\begin{itemize}
{\setlength{\rightmargin}{\leftmargin}}
\itemsep -7pt
\item{\it $v_f$ will be larger if struck on a surface with less friction because larger friction would reduce the initial momentum of the ball.}

\item{\it $v_f$ decreases with increase in $\mu$ because friction determines how much negative acceleration there is.}

\item{\it $F_f=\mu F_n$ and when $F_f$ is smaller the pool ball is impeded less and rolls faster.}

\item{\it With a smaller $\mu$ the ball will start into its roll with a greater velocity because of less opposing force to movement.}

\end{itemize}

From one on one conversations with students, it was clear that students in Category (1) often did not think through carefully about what was
causing an increase (pool ball) or a decrease (wheel) in the angular speed $\omega$ to make the object roll eventually.
However, cognitive load is typically high when individuals have to attend to several aspects of a problem
simultaneously and issues related to mental load are particularly important for those with evolving
expertise~\cite{sweller}. In fact, as noted above, even experts had difficulty thinking about both important aspects of
problem (1) simultaneously and typically focused only on one of them. 
Although students in Category (1) did not think carefully about the rotational and rolling aspects of the problem and 
focused exclusively on the linear speed, the doubling of the number in this category 
for problem (2) compared to problem (1) signifies that some students were actually carefully analyzing the problem.
The level of this analysis is commensurate with students' existing knowledge and skills and it suggests that some introductory physics students
are capable of performing sophisticated analysis of these problems.

Also, the response in Category (2) suggesting ``a higher $\mu$ implies a higher $v_f$ because friction is responsible for 
making the object roll" decreased from $27\%$ for Problem (1) to $4\%$ for Problem (2).
The reduction in the number of students in Category (2) for problem (2) compared to problem (1) also suggests that some students were
analyzing the problem deeply to the best of their ability.
The following are sample responses to problem (2) belonging to Category (2):

\begin{itemize}
{\setlength{\rightmargin}{\leftmargin}}
\itemsep -7pt
\item{\it Higher $\mu$ will give more traction to the ball and increase the $v_f$.}

\item{\it $v_f$ is larger if $\mu$ is larger because friction is the driving force for the ball to start rolling.}
\end{itemize}

Fourteen percent of the students provided responses that fell in Category (3). They believed
that higher $\mu$ implies higher $v_f$ because the pool ball will start to roll faster if the frictional
force is larger reducing the energy dissipated in the form of heat. 
Although these students focused on only one of the two important aspects of the problem, 
their responses are again reminiscent of expert responses.
The following are sample responses from Category (3):

\begin{itemize}

{\setlength{\rightmargin}{\leftmargin}}
\itemsep -7pt
\item{\it On smaller $\mu$ surface, the ball will slip for a longer time and lose more energy and move at slower $v_f$.}

\item{\it If the coefficient of friction is high, the ball will start rolling sooner before the speed is lost and its velocity will be higher.}
\end{itemize}

As in problem (1), there were responses ($6\%$) that could not be classified in any other categories
and were placed in Category (5) in Table (1).
The following is a sample response in which the student hints at friction increasing the linear velocity of the pool ball which is
the opposite of what should happen to establish rolling:

\begin{itemize}
{\setlength{\rightmargin}{\leftmargin}}
\itemsep -7pt
\item{\it $v_f$ is greater on higher $\mu$ surface because the ball will have a greater force to move forward. And the force
is related to acceleration which is related to velocity.}

\end{itemize}

\subsection{Comparison with students responding to both problems}

As noted earlier, 49 students were given both problems of the IPP. 
We wanted to understand if student response to one problem affects their response to the other problem in an IPP. 
We wanted to compare the pattern of responses for the case when both questions were asked with
the case when only one of the questions was asked of each student. 
We also discussed their responses individually with several students. 
In addition to answering both questions, students were asked an open-ended question about whether the two problems are similar or different
and they had to explain their reasoning. Although the explanation for this open-ended part was expected to vary widely,
we felt that student explanations will be useful in understanding what they viewed as the similarities or differences in the problems. 

Table (1) shows the distribution of student responses in the different categories.
The pattern of responses when both problems were given to the same student is not statistically
distinct from the case when each student answered only one of the problems. 
The comparison with the cases when students worked only on one of the problems suggests that giving both problems of the 
IPP did not give students any additional insights or make the similarity of the two problems
clear to them. As discussed earlier, many students either focused on the frictional force or the time to start rolling
and they continued to focus on the same aspects in both problems (although in
some cases they felt that the two situations will be affected differently).
The following are responses to each problem from a student who believed that a higher $\mu$ implies lower $v_f$ in
both cases:
\begin{itemize}
\item Problem (1): The friction between the wheel and the floor will cause the wheel to lose energy, which will
cause the $v_f$ to be less.
\item Problem (2): If $\mu$ is greater, the ball will have a smaller $v_f$ because of the resistance and force pushing
back against the ball. If there is no friction, the ball will roll continuously without losing energy.
\end{itemize}
The following are responses to each problem from a student who believed that a higher $\mu$ implies higher $v_f$ in
both cases:
\begin{itemize}
\item Problem (1): If $\mu$ was greater $v_f$ would be greater because energy won't be wasted as much during slipping
because it would slip for less time. So the wheel will experience less work done by the kinetic friction.
\item Problem (2): The higher $\mu$ allows for more rolling ability so greater $v_f$.
\end{itemize}
There was only one student in Category (4) who answered the problem qualitatively
correctly and provided a reasonably correct reasoning for both cases as follows:
\begin{itemize}
\item Problem (1): Since the wheel is dropped on the horizontal floor, the only [initial] velocity is in the y direction.
$v_f$ can only operate when there is friction so it will start going faster at a rougher surface but energy will
be lost quicker with more friction. On the other hand, it will take a while to get going on a surface of smaller
friction, but will lose less energy and go longer. Since energy is expelled during the slipping, no matter what the $\mu$ is the
$v_f$ is the same.
\item Problem (2): A smaller coefficient of friction will make the ball slip longer but less friction uses less energy.
The larger friction will start rolling right away but lose energy quicker. Therefore, they will have about the same speed.
\end{itemize}
In response to the open-ended question on whether the problems are similar or different, $28\%$ explicitly said that the
problems are different. For example, one student noted: ``They are different because in (1) the object is already
rotating and then is dropped to the surface whereas in (2) the object is at rest to begin with and then has
a force applied to it making it begin to move and eventually roll". Although the student focused on the difference,
the student's observation is very reasonable. The student is making a careful note of the differences in the initial conditions
of the two problems.
Another student noted: ``The wheel and ball
problems are opposite because in the wheel problem $v_f$ will be smaller if $\mu$ is smaller and in the ball
problem $v_f$ is larger if $\mu$ is smaller".
Although this student's observation is incorrect, this observation is an applaudable intuitive guess. The 
response suggests that the student has performed a conceptual analysis of the problems and realized
that friction must assist in increasing the linear speed in the first case and in decreasing the linear speed in the other
case. The following are responses to each problem from two students who believed that a higher frictional force 
will increase $v_f$ in problem (1) but not in problem (2):
\begin{itemize}
\item Problem (1): The $v_f$ is smaller if $\mu$ is smaller because some energy is lost when it is
slipping instead of rolling. $v_f$ will be smaller if it were to fall on ice because it would be sliding for
a long time than for large $\mu$ and lose more energy.
\item Problem (2): $v_f$ will be larger if $\mu$ is smaller because it [pool ball] will keep its speed as it is slipping.
\end{itemize}
\begin{itemize}
\item Problem (1): $v_f$ will be higher if $\mu$ is greater. Think about it, the wheel needs more friction
to roll rather than lose its rotational motion to slippage.
\item Problem (2): This time it is actually transferring translational movement to rotational as it slides first
and then starts to spin. A higher $\mu$ will result in a weaker slide so less $v_{cm}$.
\end{itemize}

The fact that the distribution of student response is very similar when students were given both problems in an IPP vs. only one of them
suggests that giving both problems did not significantly affect their strategy for solving each problem. As noted earlier, the correct
solution to the problems requires simultaneous focus on two variables: the magnitude of friction and the time to start rolling. 
It would have been surprising if a student could discern the importance of both of these variables to solve the problem correctly 
in one context but not in another. Such responses did not exist in our sample.
Analyzing the type of responses students provided when they were given only one of the problems in an IPP,
we cannot expect students (even those with partially correct responses) to gain additional insight about each problem when asked to 
solve both problems of the IPP.

One finding is that student expertise spans a wide range and
some student responses were better than others although not completely
correct. The implication is that it may be useful to develop objective grading schemes that account for 
different types of incorrect responses implying different levels of expertise.
Another finding is that pairing a difficult non-intuitive problem requiring proper handling of two variables 
in which the force of friction is responsible for increasing the linear speed
of the object with an isomorphic problem in which friction decreases the linear speed is unlikely to help students discern the
isomorphism between the two problems. Students had difficulty distilling the underlying physics principle involved in the problems
even when both problems were given to them at the same time compared to the case when only one of the problems was given.
A companion paper describes student performance on several IPPs with a 
range of difficulty. The goal there was to assess students' evolution of expertise and their ability to transfer from one problem
to another in an IPP over a wide variety of IPPs.

\vspace*{-.2in}
\section{Summary and Conclusion}
\vspace*{-.2in}

In this investigation, introductory physics students were given isomorphic problems related to
rotational and rolling motion that had different surface features.
We analyzed written responses to the paired non-intuitive isomorphic problems in
introductory physics courses and discussions with a subset of students about them.
We discuss why certain incorrect responses are better than others and shed light on students' evolving expertise. 
In one problem, a spinning wheel was dropped and in the other a pool ball
was struck. Students were asked to determine the role of friction in determining the final speed of the rigid wheel or pool ball
once they started to roll on the horizontal surface. The initial linear speed was non-zero in the pool ball problem but not in the other. 
Students who solved both problems in the rotational and rolling motion IPP sometimes used different knowledge resources because the initial
conditions are very different for the two problems. The fact that the linear velocity of the pool ball must decrease in order
to make it roll made the ``higher $\mu$ means lower $v_f$" idea almost twice as prevalent 
as in the spinning wheel problem where the linear speed must increase for the rolling condition to hold.
Also, roughly one fourth of the students who only solved the problem of the spinning wheel dropped to the floor thought that a higher $\mu$ 
would imply a larger $v_f$ because friction causes the wheel to roll.  
The number of students making similar claims was negligible for the pool ball problem.  
Written responses and discussions
with individual students suggest that the fact that the wheel was only spinning when it dropped on the floor forced many students to 
think that friction would help increase its linear speed. 
We believe that these students thought carefully about the problem rather than using the rote plug-and-chug strategies such
as the frictional force must always decrease the linear speed of the object. Often, students who noted that higher $\mu$
results in larger $v_f$ in the spinning wheel problem but smaller $v_f$ in the pool ball problem provided more thoughtful
although not completely correct responses than those who claimed that higher $\mu$ will always lead to smaller $v_f$.
Although students did not solve these problems correctly
(which required attention to two important variables: friction and time to start rolling), their responses in Categories (2) and
(3) in Table 1 are reminiscent of expert responses. They attest to students' evolving expertise and the fact that students were
analyzing the problems carefully commensurate with their expertise.
Some student responses were as sophisticated as those of physics professors~\cite{intuition}.
One implication for expert-novice problem solving is that 
student responses can span a wide range on the expertise scale. 
To appropriately account for students' evolving expertise, grading rubrics should be developed to favor students
who provide better responses involving deeper analysis similar to those performed by experts although the answers are incorrect. 
The grading rubric for each problem can be determined 
based upon a theoretical analysis of the problem by experts and by giving the problem to students and categorizing their responses with a focus on the quality of conceptual analysis and decision making even if the solution is incorrect.

\vspace*{-.2in}
\section{Acknowledgments}
\vspace*{-.2in}

We are grateful to J. Mestre, F. Reif, R. Glaser, H. Simon (late), 
R. P. Devaty, R. Tate, and J. Levy for useful comments and discussions.
We thank the National Science Foundation for grant NSF-DUE-0442087.

\pagebreak

\begin{center}
\epsfig{file=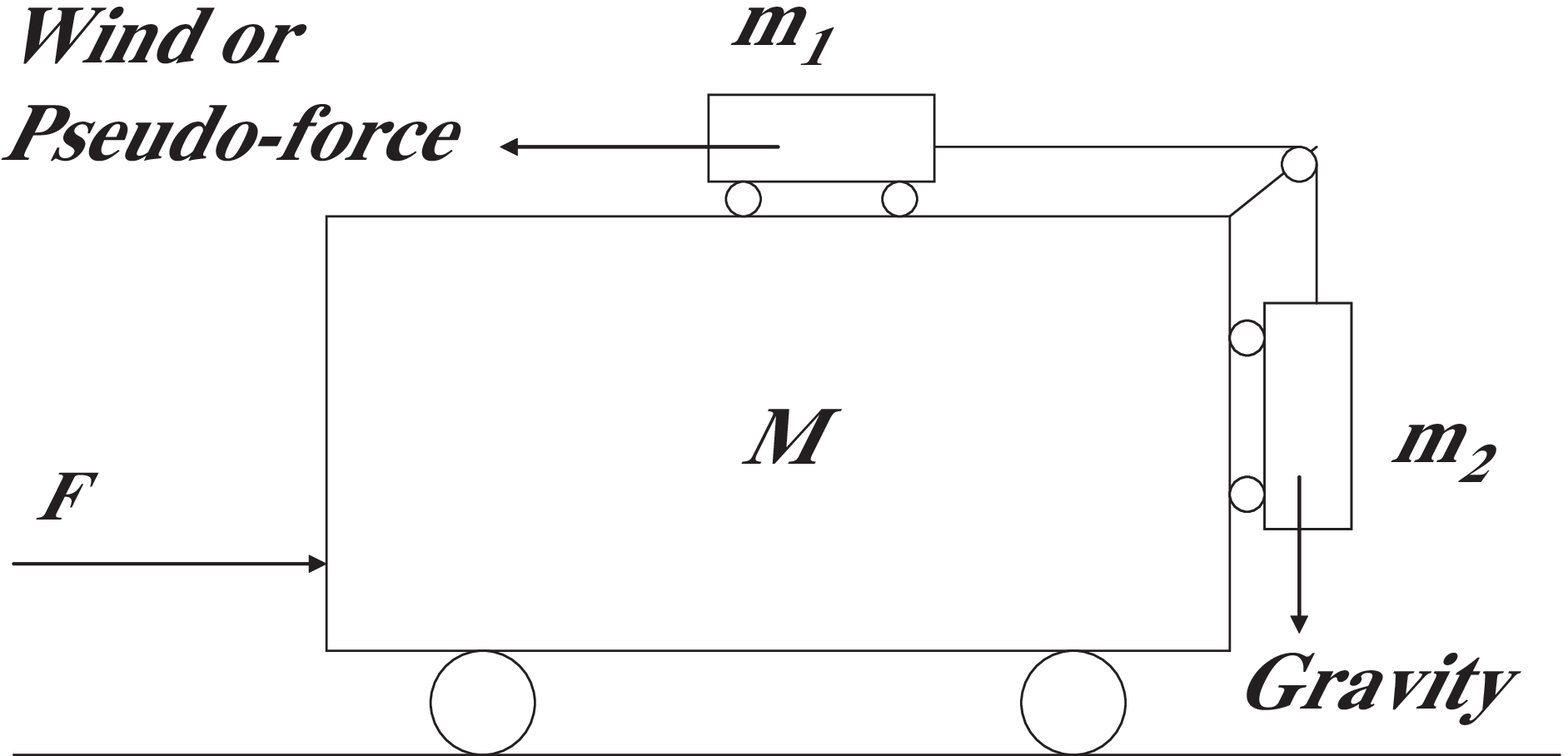,height=3.in}
\end{center}
Figure 1: Novice's and Expert's problem representation from Ref.~\cite{smith}.

\pagebreak

\begin{table}[h]
\centering
\begin{tabular}[t]{|c|c|c|c|c|}
\hline
Category& Problem (1) only&Problem (2) only&P(1) when both&P(2) when both\\[0.5 ex]
\hline \hline
1&43 &76&49&66\\[0.5 ex]
\hline
2&27 &4&27&2\\[0.5 ex]
\hline
3&9 &14&14&22 \\[0.5 ex]
\hline
4&10.5 &0&2&2\\[0.5 ex]
\hline
5&10.5 &6&8&8\\[0.5 ex]
\hline
\end{tabular}
\vspace{0.1in}
\caption{The percentage of students with responses in the five categories on the rotational and rolling motion problems. The second
and third columns are distributions of responses when students solved only problem (1) and problem (2), respectively. The fourth
and fifth columns are distributions of responses for problems (1) and (2), respectively, when the same students solved both problems.}
\label{junk}
% see table \ref{junk}
\end{table}

\end{document}